\pdfoutput=1
\documentclass[aps,prl,reprint, superscriptaddress]{revtex4-2}
\usepackage{graphicx}
\usepackage{amsmath}
\usepackage{amsfonts}
\usepackage{xcolor}
\usepackage{braket}
\usepackage{comment}
\usepackage{hyperref}
\hypersetup{colorlinks,allcolors=black}
\newcommand{\expu}[1]{\text{e}^{#1}}

\begin{document}

\title{Talbot interference of whispering gallery modes}

\author{Matias Eriksson}

\affiliation{Tampere University, Photonics Laboratory, Physics Unit, 33720 Tampere, Finland}

\author{Benjamin A. Stickler}

\affiliation{Institute for Complex Quantum Systems, Ulm University, Albert-Einstein-Allee 11, 89069 Ulm, Germany
}

\author{Robert Fickler}

\affiliation{Tampere University, Photonics Laboratory, Physics Unit, 33720 Tampere, Finland}

\date{16.7.2024}

\begin{abstract}
The Talbot self-imaging phenomenon is a fundamental interference effect that is natural to all waves with a periodic structure.  
We theoretically and experimentally study the Talbot effect for optical waves in the transverse angular domain using whispering-gallery modes of step-index multimode fibers, which carry a high orbital angular momentum and fulfill the required quadratic dispersion relation.
By using the complex interference along the core-cladding interface of these fibers, we experimentally demonstrate that the high-order fractional Talbot effect can be used to implement 9- and 30-port beamsplitters using only off-the-shelf components in a compact arrangement. 
In addition, we show that the beamsplitters can be efficiently interfaced with single-mode fibers, such that our work not only extends the recent developments on the angular Talbot effect to widely available step-index multimode fibers, but also demonstrates a powerful application as a signal multiplexer, which becomes more compact as the number of channels is increased.

\end{abstract}
\maketitle
\section{Introduction}

The Talbot effect is the self-imaging of periodic waves \cite{talbot1836lxxvi, rayleigh1881x, berry2001quantum}, which occurs when a field periodic in one domain experiences a quadratic phase modulation in its Fourier dual domain. This has been observed in many different classical and quantum waves \cite{liu1989talbot, patorski1989self, case2009realization, wen2013talbot, mcmorran2009electron, chapman1995near, nowak1997high, brezger2002matter, fein2019quantum} and their different Fourier-linked degrees of freedom, the most well-known Fourier pairs being space-momentum \cite{talbot1836lxxvi, rayleigh1881x, berry2001quantum, liu1989talbot, patorski1989self, case2009realization, wen2013talbot, mcmorran2009electron, chapman1995near, nowak1997high, brezger2002matter, fein2019quantum, AzañaJosé2014ATe} and time-frequency \cite{azana2001temporal, azana2005spectral} of optical waves. Furthermore, performing quadratic phase modulations both in a given domain and its Fourier dual gives rise to a more powerful tool called the generalized Talbot effect \cite{cortes2016generality}, capable of arbitrary control of the field periodicity in both Fourier-linked domains. Recently, the effects have been explored in the azimuthal angle- and orbital angular momentum (OAM) degrees of freedom of optical fields \cite{hu2024generalized}, where in particular the angular self-imaging effect is mediated by the quadratic dispersion of OAM modes in ring-core fibers \cite{niemeier1985self, hautakorpi2006modal, eriksson2021talbot}. The generalized angle-OAM Talbot effect can be used to control field periodicities in azimuthal angle and OAM, realize multiport beamsplitters with arbitrary numbers of ports, and perform crosstalk-free sorting of OAM modes \cite{hu2024generalized}. The last two are of particular interest in classical and quantum information processing. Beamsplitters are ubiquitous in classical and quantum information processing protocols \cite{BogaertsWim2020Ppc, flamini2018photonic}, and the OAM sorter can be used to (de)multiplex telecommunication channels in different OAM modes \cite{PuttnamBenjaminJ.2021Smfo}, or to access the infinite-dimensional OAM state space of photonic qudits \cite{ErhardManuel2018TpNq}.

This study introduces a new implementation of the angular self-imaging effect by replacing the previously used custom-tailored ring-core fibers \cite{niemeier1985self, hautakorpi2006modal, eriksson2021talbot, hu2024generalized} with a standard step-index multimode fiber (MMF). In MMFs, the self-imaging effect occurs with OAM-carrying whispering gallery modes (WGM) of the step-index fiber. The whispering gallery modes are eigenmodes of the fiber, which are of high order in the azimuthal domain (high OAM) and first order in the radial domain, accompanied by confinement to the vicinity of the core-cladding interface. 
Bearing a great resemblance to the eigenmodes of ring-core fibers, they similarly follow a quadratic dispersion curve with respect to the OAM charge, a necessary condition to realize angular self-imaging.
Although the WGM Talbot effect was among the first theoretical predictions for the realization of the angular Talbot effect \cite{baranova1998talbot}, to the best of our knowledge, our work acts as its first experimental demonstration. 
As such, we extend the angular self-imaging effect to the far more mature and readily available platform of step-index MMFs, enabling easy realization of the effect with great performance. In particular, we demonstrate the application of the WGM Talbot effect to realize multiport beamsplitters with in theory arbitrary numbers of input- and output ports.
An example of a 1-to-8 beamsplitter is shown in the conceptual Fig. \ref{fig1}a. Further, we show that our beamsplitters can be efficiently interfaced with single-mode fibers (SMF), on par with state-of-the-art multiport beamsplitter technologies.

The beamsplitters demonstrated here fall into the category of multimode interference couplers \cite{SoldanoL.B.1995Omid}, i.e., devices that employ the Talbot self-imaging effect realized through intermodal interference in waveguides to perform either $1\times N$ power splitting or $N\times N$ beamsplitting. Typically multimode interference couplers are realized in rectangular rib \cite{HosseiniA20111NMI} or strip \cite{KwongD20101×1e} waveguides, and have been demonstrated to perform low insertion loss beamsplitting with high output power uniformity for up to $1\times 12$ power splitting \cite{KwongD20101×1e}. While the number of ports of rectangular multimode interference couplers can be increased it comes at the cost of increased insertion loss and a significant decrease in power uniformity \cite{HuangJ.Z.1998Anda}, which are detrimental effects that require specialized fabrication methods to be mitigated \cite{YaoRunkui2021CaL1}.

For $1\times N$ power splitting, many competing designs for multimode interference couplers have also been realized, e.g., cascaded Y-junctions \cite{WangLiangliang2014Acal}, subwavelength grating couplers \cite{GuoZhenzhao2022Upah} and inverse tapers \cite{LiXianyao2013Cals}. Although these designs can achieve large frequency bandwidths, low insertion losses, and high output power uniformities, their fabrication complexity increases and performance decreases rapidly with increased splitting ratios $N$.

The whispering gallery Talbot effect -based beamsplitter improves on the earlier designs with it's simplicity and performance in the realm of high-order beamsplitting: The insertion losses and power uniformities are comparable to competing designs almost irrespective of the splitting ratios, i.e. for low and high $N$, as demonstrated experimentally for 9- and 30-port beamsplitters.

\section{Theoretical background}

\subsection{The azimuthal Talbot effect}

The Talbot self-imaging is the revival of periodic wave structures due to a quadratic phase modulation in the Fourier domain. The azimuthal angle is a perfect degree of freedom to realize the effect, as the field periodicity is naturally given by the periodic nature of the azimuthal coordinate itself. The azimuthal structure of a scalar field with separable radial and azimuthal components can be represented as a Fourier sum of OAM modes $\Psi(r,\varphi,z) = \sum_{\ell = -\infty}^\infty\Phi_\ell(r)\expu{\text{i}\ell\varphi}\expu{\text{i}\beta_\ell z}$,
where $(r,\varphi, z)$ are the cylindrical coordinates, and $\Phi_\ell$ is the complex amplitude of the eigenmode with topological charge $\ell$ and the associated propagation constant $\beta_\ell$. If $\beta_\ell$ has a quadratic dependence on $\ell$ of the form $\beta_\ell \propto -\pi \ell^2/z_T$ with Talbot length $z_T$, integer and fractional self-images of the initial field $\Psi(r,\varphi,0)$ are seen at integer multiples and fractions of the Talbot length, respectively. Specifically,  $q$ revivals of the initial field emerge at a propagation distance $z=p z_T/q$ with coprime $p$ and $q$, with an azimuthal shift of $2\pi p/q$ radians between individual self-images. Therefore, a field structure initially localized in the azimuthal angle can be split into multiple, azimuthally shifted copies of itself, effectively realizing a beam-splitting operation. 

\begin{figure}
    \centering
    \includegraphics[width=.48\textwidth]{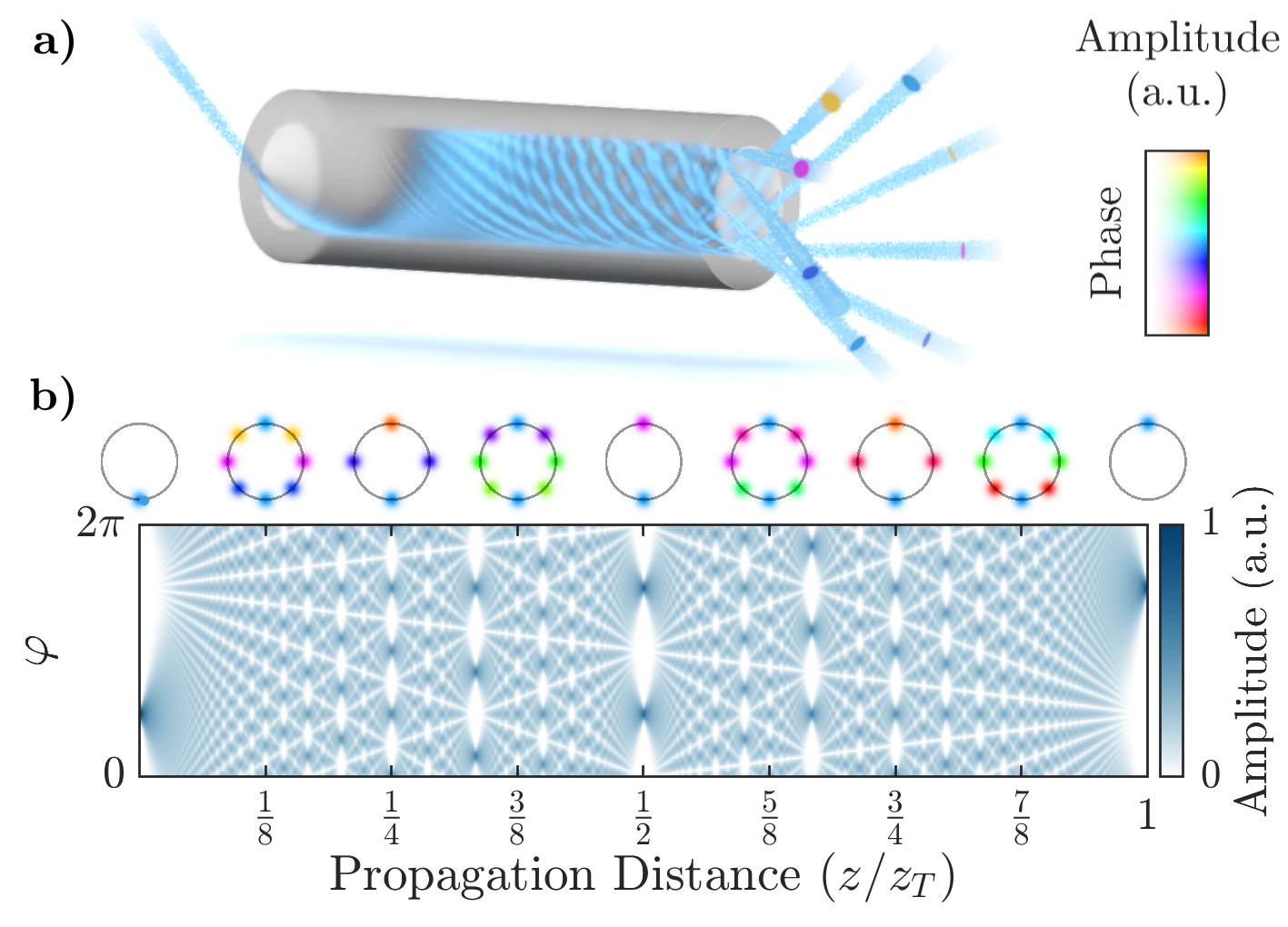}
    \caption{\textbf{a)} Conceptual figure of the whispering gallery Talbot effect in a step-index MMF. A Gaussian laser beam is coupled to the MMF with an azimuthally inclined incidence near the core-cladding interface, exciting whispering gallery modes (WGM) of the MMF. Upon propagation, the modal interference of WGMs generates a Talbot interference pattern, resulting in the emergence of an 8-fold fractional self-image of the input field at the output facet, realizing a 1-to-8 beamsplitter. \textbf{b)} Simulated propagation of an initially localized Gaussian light field in a thin cylinder shell, generating an interference pattern upon propagation known as the Talbot carpet. For visual clarity the incidence angle of the Gaussian is here set to $0$. The insets above the rolled-up carpet demonstrate the corresponding transverse amplitude distributions at different fractions of the Talbot length, showcasing the fractional and integer self-imaging of the initial field.}
    \label{fig1}
\end{figure}

This effect can be beautifully visualized by observing the associated rolled-up Talbot carpet, i.e., the azimuthal intensity evolution of an initially localized field propagating in a thin cylindrical shell, as shown in Fig. \ref{fig1}b). In this case, the Talbot length is given by $z_T = 4\pi^2R^2/\lambda$, where $\lambda$ is the wavelength and $R$ is the cylinder radius. This formula accurately predicts the Talbot length in ring-core fibers with $R$ as the mean radius of the ring-core \cite{eriksson2021talbot, hu2024generalized}, but is not directly applicable to the whispering gallery Talbot effect, where a more refined model is required.

\subsection{Modal dispersion of WGMs}

To model the Talbot self-imaging effect with the WGMs of step-index MMFs, we must first retrieve their OAM-dependent modal dispersion. WGMs are high-order OAM modes of the first radial mode order, i.e., $\text{OAM}_{\pm \ell,m}^{+}$ and $\text{OAM}_{\pm \ell,m}^{-}$ modes with OAM mode index $\ell \gg 1$ and radial mode index $m=1$. The superscript $\pm$ corresponds to left- and right-circular polarizations $\hat{\sigma}^\pm = \hat{x} \pm \text{i}\hat{y}$. 
We set $\ell$ always non-negative, and label modes with negative OAM indices by $-\ell$.

Following the well-established literature on waveguide optics \cite{SnyderA.W2012Owt}, we can calculate the form and modal dispersion of step-index MMF modes. The OAM modes of weakly-guiding step-index circular fibers with $\ell\geq2$ can be constructed as superpositions of the well-known EH and HE modes \cite{RamachandranSiddharth2013Ovif}:

\begin{subequations}
    \begin{align}
         \text{HE}^{\text{even}}_{\ell+1,m} \pm \text{i}\text{HE}^{\text{odd}}_{\ell+1,m} = \text{OAM}_{\pm \ell,m}^{\pm} \label{eq:HE_superpos}\\
         \text{EH}^{\text{even}}_{\ell-1,m} \pm \text{i}\text{EH}^{\text{odd}}_{\ell-1,m} = \text{OAM}_{\pm \ell,m}^{\mp} \label{eq:EH_superpos}
    \end{align}
\end{subequations}

Specifically, the electric field components of the OAM modes can be written as:

 \begin{align}
         \text{OAM}_{\pm \ell,m}^{s}:
         \begin{cases}
             \boldsymbol{e_\bot} =  \hat{\sigma}^s F_{\ell,m}(r)\expu{\pm\text{i}\ell\varphi}\\
             e_z = \frac{\text{i}}{k_{\text{co}}R}G^{\mp s}_{\ell,m}(r)\expu{\pm\text{i}(\ell\pm s)\varphi}
         \end{cases}
 \end{align}
 where $s =\pm 1$ and $k_\text{co}(k_\text{cl})$ is the wavenumber in the core (cladding) and $R$ is the core radius. With modal parameters $U = R\sqrt{k_{\text{co}}^2 - \beta_{\ell,m}^2}$ and $W = R\sqrt{\beta_{\ell,m}^2 - k_{\text{cl}}^2}$, the explicit forms of $F_{\ell,m}$ and $G_{\ell,m}^{\pm}$ are

\begin{subequations}
\begin{align}
    F_{\ell,m}(r) = 
    &\begin{cases}
        \frac{1}{J_\ell(U)}J_\ell\left (\frac{Ur}{R} \right )  \hspace{28pt}&\hspace{2.5pt}0\leq r \leq R \\
        \frac{1}{K_\ell(W)} K_\ell\left (\frac{Wr}{R}\right )  &R\leq r < \infty
    \end{cases}  \\
    G_{\ell,m}^{\pm}(r) = 
    &\begin{cases}
        \pm \frac{U }{J_\ell(U)} J_{\ell\mp 1}\left (\frac{Ur}{R} \right )&\hspace{2.5pt}0\leq r \leq R \\
        \pm \frac{W}{K_\ell(W)} K_{\ell\mp 1}\left (\frac{Wr}{R}\right )&R\leq r < \infty
    \end{cases} 
\end{align}
\end{subequations}
where $J_\ell$ is the Bessel function of the first kind and $K_\ell$ is the modified Bessel function of the second kind.

For slightly radiative, or leaky modes, with $\mathrm{Re}(\beta)$ slightly below $k_{\text{cl}}$ and $\mathrm{Im}(\beta)\neq 0$, it is customary to explicitly note their radiative nature by replacing the cladding mode parameter with $Q=\text{i}W$ and the $K_\ell$ Bessel function to the non-evanescent Hankel function of the first kind $K_\ell(x) = \frac{\pi}{2}\text{i}^{\ell+1}H_\ell^{(1)}(\text{i}x)$, however both forms are functionally equivalent. These leaky modes can be useful as they are naturally resilient to intermodal coupling and can propagate for long distances with negligible loss \cite{ma2023scaling}.

From the core-cladding interface conditions for the fields we can derive the following eigenvalue equation
\begin{equation} \label{eq:bound_eigenvalues}
    U \frac{J_{\ell+1}(U)}{J_\ell(U)} = W\frac{K_{\ell+1}(W)}{K_\ell(W)} \equiv Q\frac{H_{\ell+1}^{(1)}(Q)}{H_\ell^{(1)}(Q)}
\end{equation}

\begin{figure} 
    \centering
    \includegraphics[width=.48\textwidth]{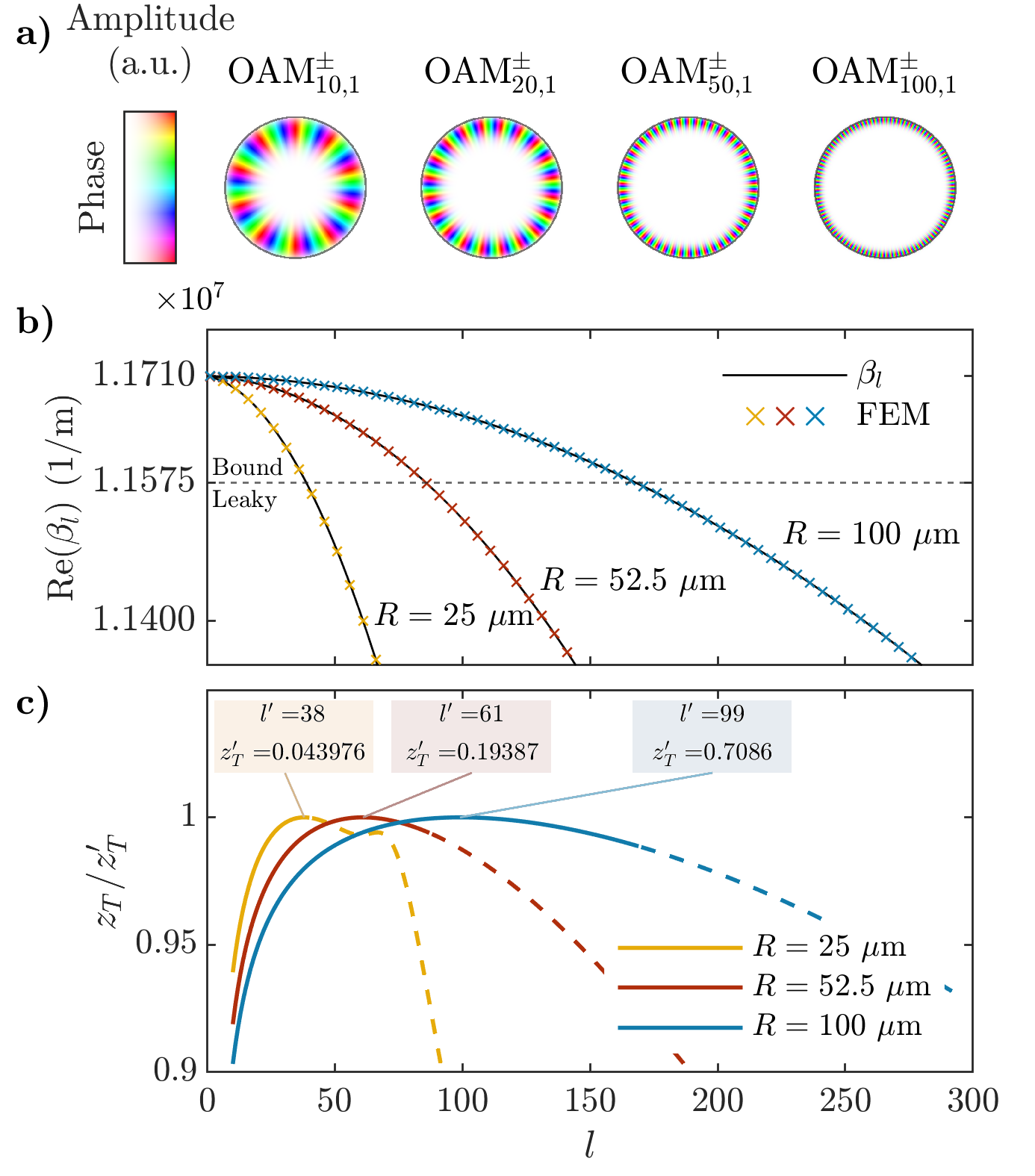}
    \caption{Whispering gallery modes of the step-index MMF. \textbf{a)}~Transverse fields $\boldsymbol{e_\bot}$ of WGMs, i.e., $\text{OAM}_{\ell,m=1}^{\pm}$ modes of a MMF with $R=52.5~\mu$m. The core-cladding interface is shown by the gray circles. 
    \textbf{b)}~OAM dispersion of WGMs for MMFs with core radii $25~\mu$m, $52.5~\mu$m and $100~\mu$m. The solid line corresponds to the propagation constants calculated from Eq. (\ref{eq:bound_eigenvalues}), and the crosses correspond to the same propagation constants calculated with finite element method (FEM) in COMSOL. The crosses are plotted only for every fifth index $\ell$ for clarity. The differences in $\beta_\ell$ between the mode groups for a given $\ell$ are not visible in the figure. The dashed horizontal line at $\mathrm{Re}(\beta_\ell)=k_{\text{cl}}$ corresponds to the boundary between bound and leaky mode regimes. 
    \textbf{c)}~OAM dependent Talbot length. Talbot lengths calculated from polynomial fits to $\mathrm{Re}(\beta_\ell)$ in ranges $[\ell-5,\ell+5]$ as described in the main text. Relative Talbot lengths $z_T/z_T'$ are obtained by dividing by the peak Talbot length $z_T'$ found around $\ell'$ for all core sizes. The dashed part of the lines correspond to the leaky mode regimes for each fiber.
    } 
    \label{fig2}
\end{figure}

This equation is solved numerically, and the solutions fix the propagation constants $\beta_{\ell,m}$ of the eigenmodes of the MMF for OAM indices $\pm \ell$, therefore also defining the exact forms of all field components. From the propagation constants that solve Eq. (\ref{eq:bound_eigenvalues}) for a given $\ell \gg 1$, the one with the highest-valued real part corresponds to the propagation constant of the WGM, i.e., $\beta_{\ell,m=1}$. The transverse fields of select WGMS are shown in Fig. \ref{fig2}a.

Generally the four modes for a given $\ell\geq2$ are split into two degenerate pairs based on the mutual alignment of the circular polarization ${\pm}$ and the OAM index $\pm \ell$, such as in Eqs (\ref{eq:HE_superpos}) and (\ref{eq:EH_superpos}). The weakly-guiding approximation neglects any polarization effects rendering all four modes degenerate.
Although in a more accurate picture small polarization corrections should be applied, they are negligible in our experiment and thus not considered here (see supplementary material).

To show the quadratic OAM dispersion of the WGMs, we plot $\beta_{\ell,m=1}$ solved from Eq. (\ref{eq:bound_eigenvalues}) as a function of the OAM mode index $\ell$ in Fig. \ref{fig2}b. To verify the accuracy of the weak-guiding approximation, we overlay the plot with propagation constants simulated with a finite element method in COMSOL. The parameters used are $\lambda_0 = 780~$nm, $n_{\text{co}} = 1.4537$, $n_{\text{cl}} = \sqrt{n_{\text{co}}^2-\text{NA}^2} = 1.4370$ with numerical aperture $\text{NA}=0.22$. The core refractive index corresponds to the refractive index of SiO\textsubscript{2} at a vacuum wavelength of 780 nm \cite{MalitsonI.H1965ICot}. The cladding and core diameters are set to $d=125~\mu$m for $R=25~\mu$m and $R=52.5~\mu$m, and $d=220~\mu$m for $R=100~\mu$m.

In the theoretical section we concluded that the azimuthal Talbot effect is seen with a quadratic OAM dispersion of the form $\beta_\ell \propto -\pi \ell^2/z_T$. The Talbot lengths for the different fibers can be estimated by fitting 2nd-order polynomials $\beta_\ell=a_\ell \ell^2+b_\ell \ell+c_\ell$ to the dispersion curves. Since higher-order dispersion terms are generally not negligible over the entire dispersion curve, the quadratic fits must be performed locally to $\beta_\ell$ in small OAM intervals. The $a_\ell$-term of the fit determines the local curvature, and therefore the local, OAM-dependent Talbot length as $z_{T,\ell} = -\pi/a_\ell$. The remaining terms are largely irrelevant in terms of self-imaging: $b_\ell$ constitutes a rotation of the field upon propagation, while $c_\ell$ contributes an irrelevant global phase across the WGMs. 
While the linear and constant terms vary freely over $\ell$, the quadratic terms alone accurately model the OAM-dependent Talbot length for fields sufficiently localized in $\ell$ to a regime where the fit is valid (see supplementary material). Similar results to the quadratic terms of the local fits can be obtained by differentiating twice a single very high-order polynomial fitted to the entire dispersion curve, converging to the same curve as the order of the polynomial is increased.

We perform the local fits to each of the dispersion curves in intervals of $[\ell-5, \ell+5]$ for $\ell \geq 10$ with coefficients of determination R-squared $>1-10^{-6}$. From the fits we retrieve+ the relative OAM-dependent Talbot lengths for the three fibers plotted in Fig. \ref{fig2}c, normalized by the peak Talbot lengths retrieved from each dispersion curve for easier comparison of the different fibers.
The Talbot lengths for the three fibers are peaked at $\ell'=38, 61$ and $99$, with peak Talbot lengths of $z_T'=43.9~$mm, $193.9~$mm and $708.6~$mm.
While $z_T\propto \beta_\ell''$, the third derivative of $\beta_\ell$ is near zero around the peaks, and the modal dispersion is closest to quadratic. As the Talbot self-imaging is an effect of purely quadratic dispersion, the peaks mark the regions of the most optimal OAM for realizations of the whispering gallery Talbot effect.
In other words, as the Talbot length is nearly constant around the peaks, the highest number of neighboring WGMS which can mutually contribute to the Talbot self-imaging without significant dephasing from higher-order terms of $\beta_\ell$ are seen in the peak regions.

\section{Experiment}

\subsection{The experimental setup}

\begin{figure}
    \centering
    \includegraphics[width=0.48\textwidth]{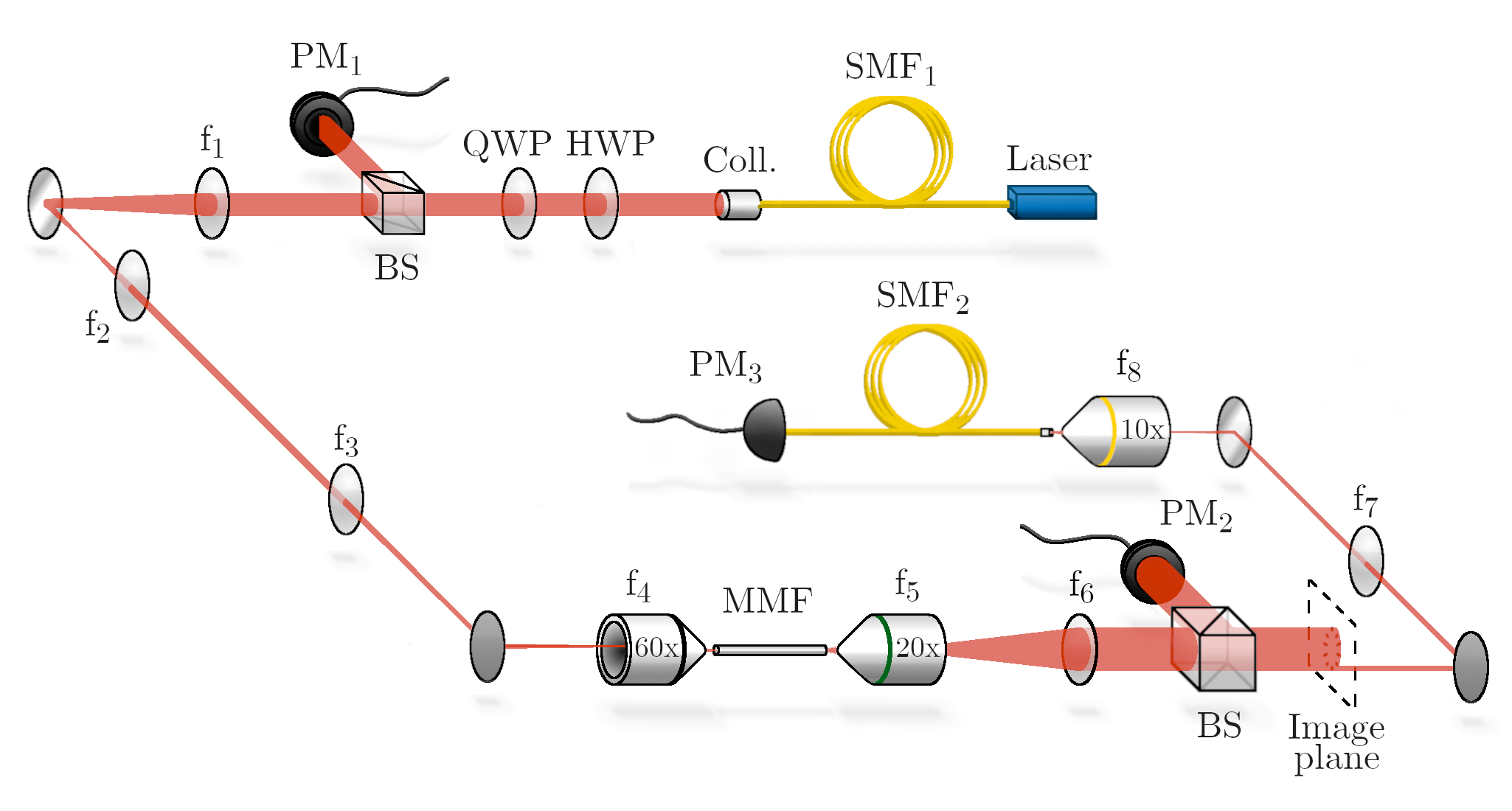}
    \caption{The experimental setup. SMF = Single-mode fiber. Coll. = Collimator. QWP/HWP = Quarter-/Half-waveplate. BS = beamsplitter. PM = Powermeter. MMF = Multimode fiber. In the image plane shown with the dashed square, either a camera or an opaque mask is placed, to capture images of the MMF end facet or to transmit a single output spot to couple to SMF$_2$.}
    \label{fig3}
\end{figure}

The experimental setup, shown in Fig. \ref{fig3}, consists of a wavelength-tunable CW laser (Toptica DL Pro, $\lambda\approx$ 756-815~nm) and imaging optics to focus the laser beam onto the input facet of the MMF, to image the output facet of the MMF and to couple a single output port of the implemented beamsplitter to an SMF.
The laser light coupled to an SMF was first collimated with a measured beam waist radius of $1.03~$mm, and demagnified onto the fiber input facet via two cascaded 4f-systems (lenses with $f_{1,2,3} = 200/150,50,300~$mm and a $60\times$ microscope objective with an effective focal length $f_4 = 2.67~$mm). The beam is focused on the fiber facet with an azimuthally inclined incidence near the core-cladding interface, exciting a spectrum of WGMs with a Gaussian OAM spectral shape \cite{Yao:06}. The reported waist radii are simply estimates, and further optimized during alignment for optimal coupling by adjusting the placement of the imaging optics. As the beam waist is similar to the radial width of the WGMs, the radial input position and azimuthal incidence angle can be tuned in parallel to minimize the excitation of unwanted higher radial order modes, and to optimize the mean value of the excited OAM spectrum for the best self-imaging conditions. A more detailed analysis on coupling a Gaussian beam to the WGMs of a MMF can be found in the supplementary material.

The short fiber pieces were laid on a metallic plate without mechanical fixation to avoid small deformations of the fiber, which would lead to unwanted mode mixing, degrading the self-imaging process. The fibers are short enough to
remain perfectly straight when laid freely on a flat surface, enabling coupling-free propagation of the fiber eigenmodes.

The output facet of the MMF was imaged using a subsequent 4f-system ($20\times$ microscope objective with $f_5=8~$mm and a lens $f_6 = 150~$mm). In the image plane, we placed either a camera to capture images of the output light field, or an opaque mask that transmits one of the output spots. With a final 4f-system (lens $f_7 = 300~$mm and $10\times$ microscope objective with $f_8=16~$mm), the transmitted output spot was coupled to an SMF connected to a powermeter, where the SMF can be considered as a single output channel of the multiport beamsplitter device. Along the beam path we placed beamsplitters and power meters to measure the power of the light field at different points along the setup, used to calculate the coupling efficiencies between the SMF input- and output channels and the intermediate MMF.

\subsection{Whispering gallery self-imaging}

To demonstrate the self-imaging phenomenon we used two pieces of standard step-index MMFs with a pure SiO$_2$ core, $\text{NA} = 0.22$. The two fibers have core radii $52.5~\mu$m and $100~\mu$m, and the peak Talbot lengths $z_{T1}=19.38~$cm and $z_{T2}=70.86~$cm, respectively (see Fig. \ref{fig2}c). The former is cut to an approximate length of $z_{T1}/9 = 2.15$~cm for 9-fold fractional self-imaging, and the latter to $z_{T2}/30 = 2.36$~cm for 30-fold fractional self-imaging. While our cleaving method is not precise enough to cleave the fiber at exactly a fraction of the Talbot length, we make use of the wavelength-dependence of the Talbot length $(z_T\propto \lambda^{-1})$: By slightly tuning the wavelength of our laser, we can adjust the self-imaging length to compensate for the imprecise cleaving.

To excite WGMs of high OAM, the incident laser beam is focused onto a small spot ($w_{\text{1-to-9}} \approx 2.3\mu$m, $w_{\text{1-to-30}} \approx 3.1\mu$m, further optimized in alignment) near the core-cladding interface with an azimuthally inclined incidence angle. 
Via the Talbot interference phenomenon the input light field is self-imaged at the output facets of the MMFs, leading to 9 and 30 azimuthally rotated copies of the incident input spot, as shown in the camera images in Fig. \ref{fig4}.

\subsection{Multiport beamsplitters}

The operation of splitting the single input spot into multiple copies of itself can be understood as a beamsplitting operation, i.e., the 9- and 30-fold fractional self-imaging systems can be thought of as 9-to-9 and 30-to-30 beamsplitters with one input port illuminated.
While we only illuminate the fiber with a single input beam and perform 1-to-N beamsplitting, the other N-1 input ports would be accessible by similar input beams azimuthally rotated by $2\pi n/N$, $n\in \mathbb{Z}$. In this way, the fractional self-images of the different input beams would be perfectly overlapped at the output plane, realizing an N-to-N coupler. For a quantitative analysis of the implemented multiport beamsplitters, we calculated the power splitting ratios between the output ports, and measured coupling efficiencies between the input- and output SMFs and the MMF.

\begin{figure}
    \centering
    \includegraphics[width=0.48\textwidth]{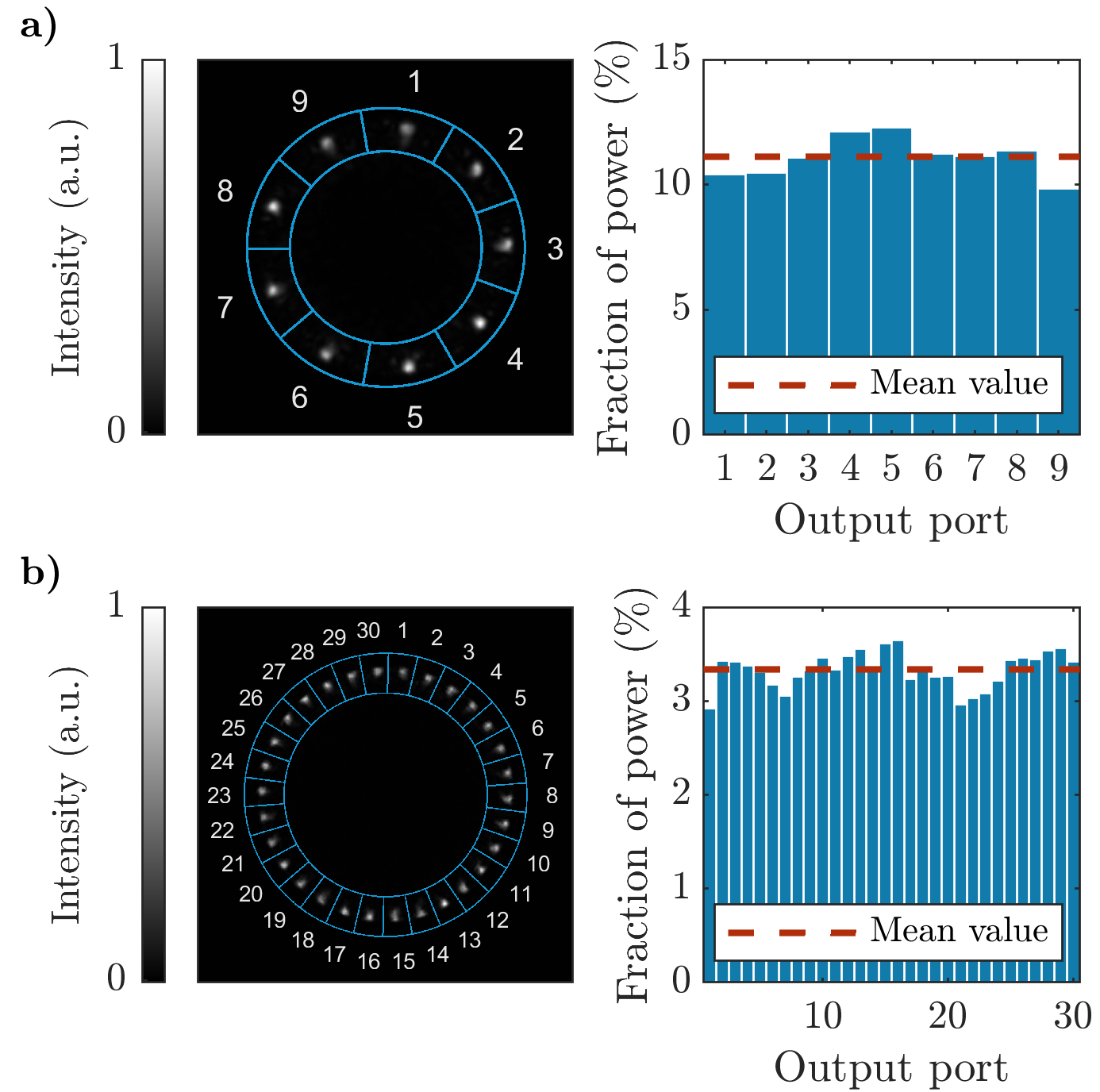}
    \caption{The output facet of \textbf{a)} 1-to-9 and \textbf{b)} 1-to-30 beamsplitters employing the whispering gallery Talbot effect with one input port illuminated, i.e., the 9- and 30-fold fractional self-images of the input field. Left: Images of the output facets of the MMFs used to realize the multiport beamsplitters. Right: The relative power distribution across all output ports, calculated through integrating the pixel values of the sectors designated in the images on the left.}
    \label{fig4}
\end{figure}

The relative power distributions between the different output ports were calculated using the camera images in Fig. \ref{fig4}. First, a background intensity value, or the mean pixel value of a dark area of the camera sensor was subtracted from each pixel. The relative power distribution across all self-images was then retrieved by comparing the sums of the pixel values over regions associated with each self-image.
From the power distributions we can determine the output uniformities as $U = -10\text{log}[\text{min}(P_i)/\text{max}(P_i)] = $ 0.95~dB and 0.97~dB for the 1-to-9 and 1-to-30 beamsplitters, respectively.

In order to test the efficiency of interfacing the WGM beamsplitter with SMF input- and output channels, we blocked all but one output spot with an opaque mask, coupled the transmitted spot to an SMF, and measured the power level of the laser field at different positions along the beam line. Three power meters (see Fig. \ref{fig3}) were used to measure the input power $P_1$, the power of the total field after the MMF $P_2$, and power of a single output port coupled to an SMF$_2$ $P_3$. All power values were taken as the means and standard deviations of $>1000$ consecutive power measurement points. From the power measurements, we can calculate the input coupling efficiency of the beamsplitter as $\eta_\text{i}=P_2/P_1$, and the output coupling efficiency as $\eta_\text{o}=P_3/P_2'$, where the reference power $P_2'$ is a fraction of the total power after the MMF associated with the chosen output port (see Fig. \ref{fig4}).
In addition, we characterize and account for the reflectance and transmittance of the beamsplitter cubes as well as the loss of the microscope objectives which lack the appropriate AR-coating. 

For the 1-to-9 beamsplitter we measured the input (SMF$_1$-MMF) and output (MMF-SMF$_2$) coupling efficiencies $\eta_\text{i} = 55.39 \pm 0.09\%$ and $\eta_\text{o} = 61.11\pm 0.50\%$, where the latter measurement was done with port 3 in Fig. \ref{fig4}a coupled to the output SMF$_2$.
In terms of coupling losses in dB, this equates to $\text{L}_{\text{i}}=-10\times\text{log}(\eta_\text{i}) = 2.57 \pm 0.01$~dB and $\text{L}_{\text{o}}=-10\times\text{log}(\eta_\text{o}) = 2.14 \pm 0.04$~dB.
Similarly for the 1-to-30 beamsplitter, we measured $\eta_\text{i}=66.40\pm0.12\%$ ($\text{L}_{\text{i}} = 1.78 \pm 0.01$~dB), and using output port 9 in Fig. \ref{fig4}, we obtained $\eta_\text{o}=66.33\pm0.19\%$ ($\text{L}_{\text{o}} = 1.78 \pm 0.01$~dB).

Assuming that the output coupling efficiencies of the chosen ports are representative of all output ports for each beamsplitter, we can estimate the total beamsplitter efficiencies as $\eta_{\text{tot}} = \eta_\text{i}\eta_\text{o} = 34.18\pm0.29\%$ and $44.04\pm0.15\%$ for the 1-to-9 and 1-to-30 beamsplitters, respectively. This equates to insertion losses of $\text{IL} = -10\times\text{log}(\eta_\text{i}\eta_\text{o}) = \text{L}_{\text{in}}+\text{L}_{\text{out}} = 4.70 \pm 0.04$~dB and $3.55 \pm 0.01$~dB.

\section{Conclusion}

We performed the first experimental study of the whispering gallery mode Talbot effect in standard step-index MMFs, after it's theoretical prediction already over 25 years ago \cite{baranova1998talbot}.
Our results expand the azimuthal self-imaging to a new platform making the effect easily realizable with standard, off-the-shelf fibers. 
We further demonstrate that the WGM Talbot effect can be used as a powerful multiport beamsplitter especially competitive in high-order beamsplitting, with the additional feature of becoming more compact the larger the splitting ratio for a given fiber. Although we achieved competitive values in terms of uniformity and efficiency, both can certainly be improved with a more elaborate assembly and alignment, or ultimately in an entirely integrated setting with no free-space optical elements.

With its performance and ease of application, the whispering gallery Talbot effect could find much use as a multiport beamsplitter in classical and quantum photonic networks \cite{BogaertsWim2020Ppc, flamini2018photonic}. The effect can further be used to realize a theoretically crosstalk-free OAM sorting scheme \cite{hu2024generalized}, useful in OAM (de)multiplexing of telecommunication channels \cite{PuttnamBenjaminJ.2021Smfo} and to access the infinite-dimensional OAM state space of photons for quantum information processing \cite{ErhardManuel2018TpNq}.

\begin{acknowledgments}
This work was supported by the European Research Council (TWISTION, 101042368). ME acknowledges the Research Council of Finland Flagship Programme, Photonics Research and Innovation (PREIN), 320165. RF acknowledges the Research Council of Finland through the Academy Research Fellowship (Decision 332399). BAS acknowledges funding by the DFG--510794108 as well as by the Carl-Zeiss-Foundation through the project QPhoton. ME, BAS, and RF declare a pending European patent application (23178785.4). The data underlying the results presented in this paper may be obtained from the authors upon reasonable request.
\end{acknowledgments}

\bibliography{main}

\end{document}


\title{Supplementary information for: Talbot interference of whispering gallery modes}

\author{Matias Eriksson}

\affiliation{Tampere University, Photonics Laboratory, Physics Unit, 33720 Tampere, Finland}

\author{Benjamin A. Stickler}

\affiliation{Institute for Complex Quantum Systems, Ulm University, Albert-Einstein-Allee 11, 89069 Ulm, Germany
}

\author{Robert Fickler}

\affiliation{Tampere University, Photonics Laboratory, Physics Unit, 33720 Tampere, Finland}

\date{16.7.2024}
\maketitle

\section{Azimuthal Talbot self-imaging}

The azimuthal structure of a scalar light field can be represented as a Fourier sum of OAM modes $\Psi(r,\varphi,z) = \sum_l \Phi_l(r,z)\expu{\text{i}l\varphi}$,
where $(r,\varphi, z)$ are the cylindrical coordinates, and $\Phi_l$ is the complex amplitude associated with the OAM mode with the topological charge $l$. Assuming the field is single-moded in the radial degree of freedom, the propagation of $\Phi_l$ is reduced to a simple propagation-induced phase factor with an associated propagation constant $\beta_l$:

\begin{equation}
    \Psi(r,\varphi,z) = \sum_l \Phi_l(r)\expu{\text{i}l\varphi}\expu{\text{i}\beta_l z}
\end{equation}

Now, given that $\beta_l$ has a quadratic dependence on $l$, Talbot self-imaging can occur. With $\beta_l \propto -\pi l^2/z_T$ where $z_T$ is the Talbot length, integer and fractional self-images of the initial field $\Psi(r,\varphi,z=0) = \Psi_0(\varphi)$ are seen at propagation distances of integer multiples and fractions of $z_T$, respectively:
\begin{align}
        \Psi(\varphi,z &= z_T) = \Psi_0(\varphi+\pi) \\
        \Psi(\varphi,z &= z_T/2) = \frac{\expu{-\text{i}\pi/4}}{\sqrt{2}}(\Psi_0(\varphi) + \text{i}\Psi_0(\varphi+\pi))\\
        \Psi(\varphi,z &= z_T/3) = \frac{1}{\sqrt(3)}(\expu{\text{i}\pi/6}\Psi_0(\varphi+2\pi/3) - i\Psi_0(\varphi) \nonumber \\ & \hspace{50pt}+ \expu{\text{i}\pi/6}\Psi_0(\varphi+4\pi/3))
\end{align}

\section{OAM-dependence of whispering gallery Talbot self-imaging}

As a demonstration of the OAM-dependent Talbot length model, we simulate the propagation of fields generated as superpositions of WGMs with a Gaussian-shaped OAM spectra with varying mean OAM values. Initially, the superposition of the WGMs generates a localized spot, similar to a Gaussian laser beam incident on the fiber with at an azimuthally inclined angle. According to the OAM-dependent Talbot length model (see Fig. 3c of the main text), the fields excited by different incidence angles (corresponding to OAM spectra with different mean OAM) are self-imaged at slightly different propagation distances. To show this, we simulate the propagation of these fields by applying propagation-induced phase terms to the individual WGMs according to their propagation constants $\beta_l$, as $\Psi(z) = \sum_l \text{OAM}_{l,1}\times\expu{\text{i}\beta_l z}$, where the propagation constants $\beta_l$ are retrieved from Eq. 4 in the main text. 

\begin{figure}
    \centering
    \includegraphics[width=.48\textwidth]{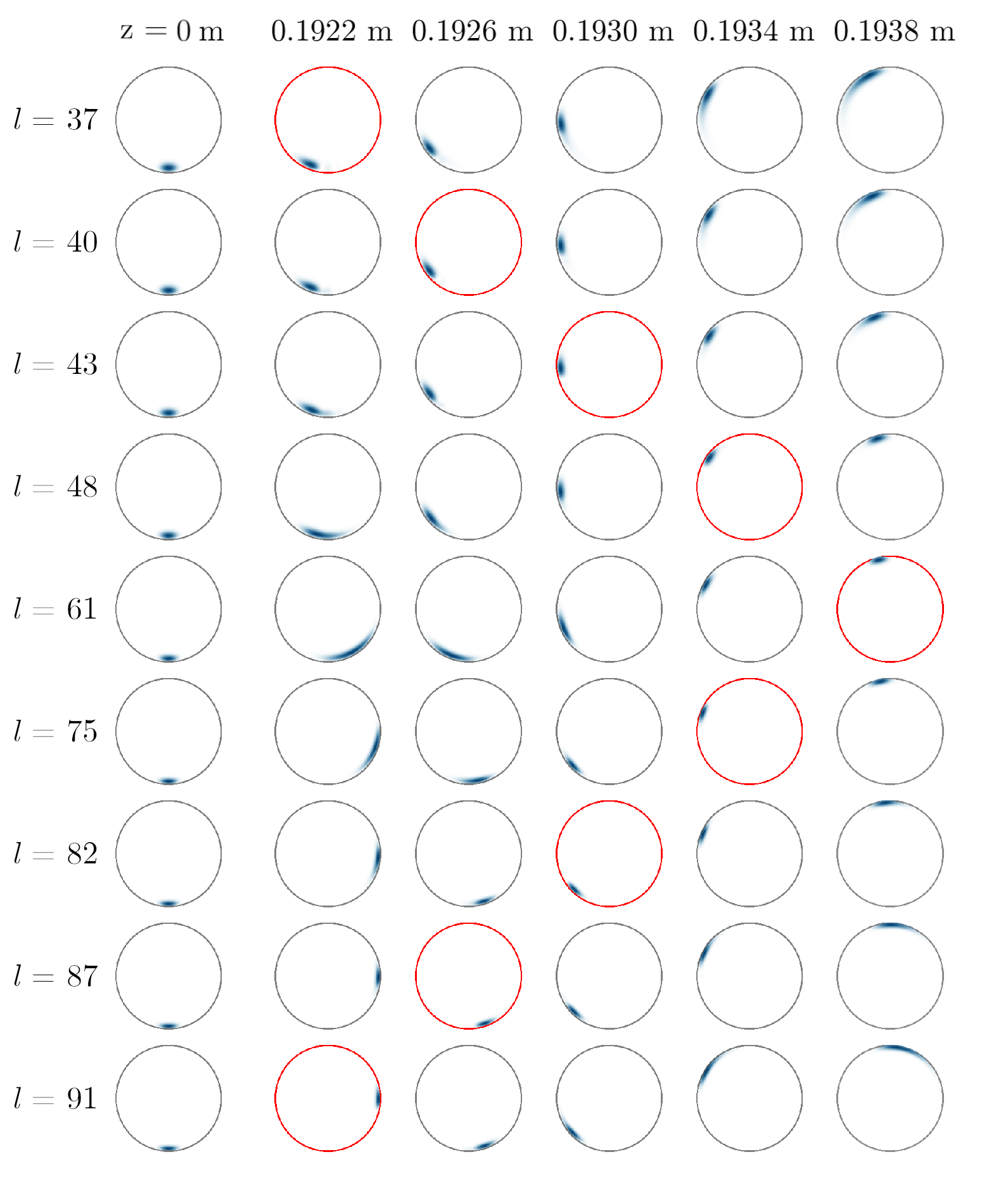}
    \caption{OAM-dependent Talbot length. Simulated evolution of fields generated as superpositions of WGMs with Gaussian OAM spectra with different mean OAM $l$, corresponding to initially localized spots with different incidence angles (at $z=0$).
    The different mean OAM charges $l$ lead to different Talbot lengths as shown in Fig. 2c in the main text. The propagation distances corresponding to the Talbot lengths for each mean $l$ are highlighted in red. The rotation of the self-images is due to the linear component of the modal dispersion $\beta(l)$. Without this linear contribution, this self-image would be seen at the top of the fiber, exactly opposite from the input spot.}
    \label{fig_s1}
\end{figure}

In Fig. \ref{fig_s1} we plot fields with different mean OAM propagated at a few distances near the Talbot length, corresponding to the OAM-dependent Talbot lengths for the different mean OAM values. Here, we observe that not only are the different fields self-imaged at different distances (accurately represented by the model in Fig. 2c in the main text), we further see an OAM-dependent rotation of the self-images due to the linear terms of the modal dispersion.

\section{Coupling Gaussian beams and WGMs}

\begin{figure}
     \centering
     \includegraphics[width=.48\textwidth]{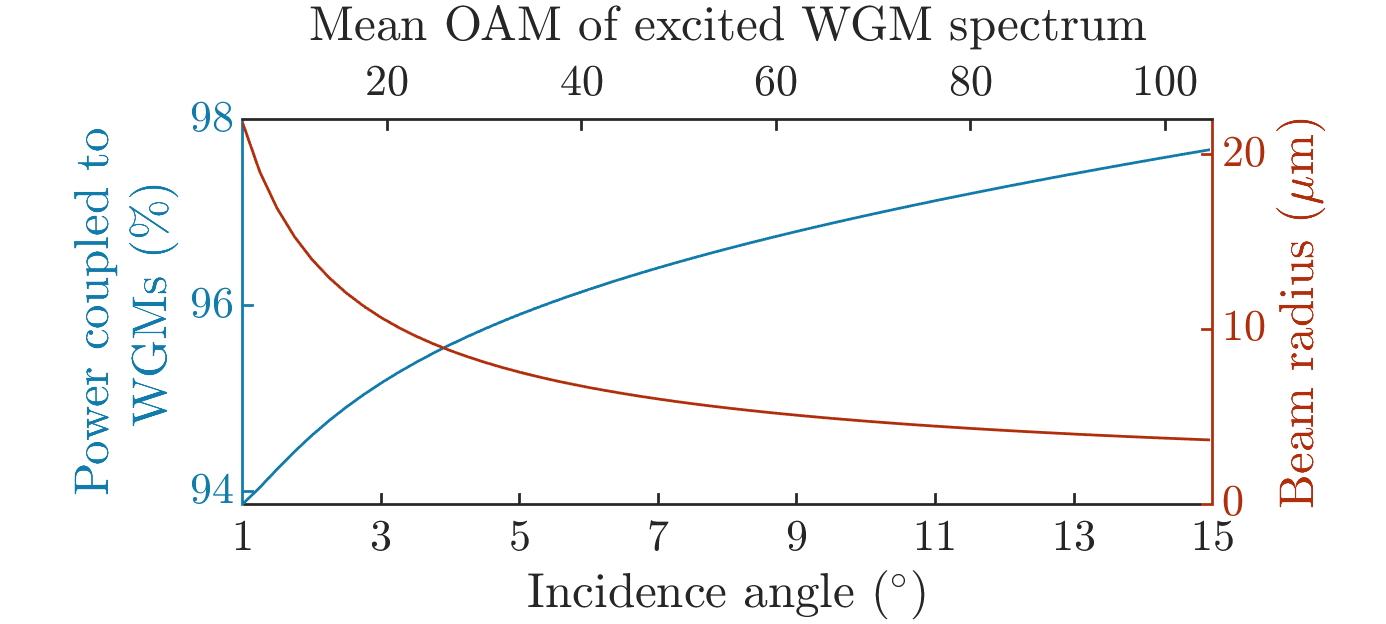}
     \caption{Optimized coupling of azimuthally tilted Gaussian beams to WGMs of a step-index MMF with a core radius of $52.5~\mu$m. Blue: relative power coupled to WGMs for Gaussian input beam optimized for coupling to WGMs. Red: The $1/\text{e}^2$ beam radius $w$ of the optimized Gaussian input beam.}
     \label{fig_s2}
 \end{figure}

Focused Gaussian beams incident on step-index MMFs with an azimuthal incidence angle $\theta$ near the core-cladding can be optimized for rather efficient coupling to WGMs. To demonstrate this, we optimize the parameters of azimuthally tilted Gaussian beams $\Psi = \text{exp}(-2((x-x_0)^2+(y-y_0)^2)/w^2)\times\text{exp}(\text{i}k_0 x \times \text{sin}(\theta))$ with $\theta \in [1, 15]$ for maximal coupling to WGMs of a step-index MMF with core radius $R=52.5~\mu$m. The optimization of the parameters is done via a Nelder-mead algorithm. Other parameters are set according to the main text. The relative powers of the excited OAM modes of the fiber are calculated through overlap integrals $|a_{\pm l,m}|^2 = \left|\int \Psi^*\text{OAM}_{\pm l,m}/|\Psi||\text{OAM}_{\pm l,m}|\right|^2$.
 We plot the ratio of power coupled to WGMs, as well as the optimized spot size as a function of the azimuthal incidence angle (or equivalently the mean OAM of the excited WGM spectrum) in Fig. \ref{fig_s2}.

\begin{figure}
     \centering
     \includegraphics[width=.48\textwidth]{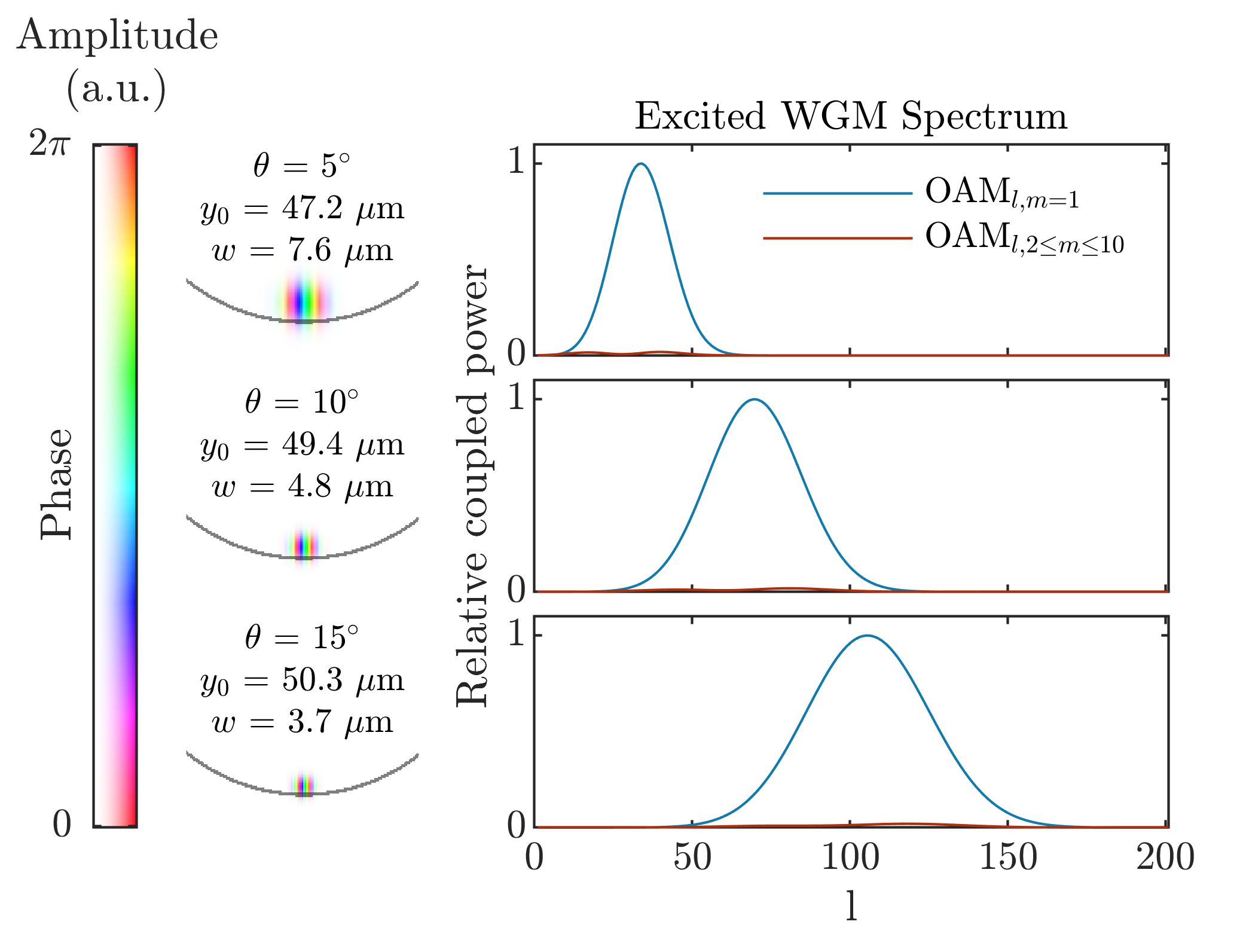}
     \caption{Horizontally tilted Gaussian input beams optimized for coupling to WGMs of a step-index MMF with a $52.5~\mu$m core radius. Left: Complex fields of Gaussian beams horizontally tilted by $\theta \in [5^\circ,10^\circ,15^\circ]$ with $x_0=0$, and vertical input positions $y_0$ and $1/e^2$ beam radii $w$ optimized for coupling to WGMs of the fiber. The gray arc shows a part of the core-cladding interface. Right: The excited WGM power spectrum of the incident beam, as well as the power coupled to higher radial order modes up to order $m=10$.}
     \label{fig_s3}
\end{figure}

 In Fig. \ref{fig_s3} we show a few Gaussian beams with azimuthal incidence angles $\theta \in [5^\circ,10^\circ,15^\circ]$ optimized for coupling to WGMs, as well as the WGM power spectra excited by the optimized beams. The optimized beam radius and input position are determined by the radial width of the WGMs, which shrinks with increasing OAM. Since we keep the Gaussian beams symmetric in the horizontal and vertical (equiv. azimuthal and radial) degrees of freedom, the beams optimized for higher OAM modes are necessarily smaller also in the azimuthal direction, leading to wider excited OAM spectra due to the Fourier duality between azimuthal angle and OAM \cite{Yao:06}. However, this symmetry between the two transverse degrees of freedom of the incident beam is by no means necessary for optimal coupling to WGMs, and can be broken to adjust the excited WGM spectra at will.

\section{Error of the weakly-guiding approximation and the scalar propagation constant}

The weakly-guiding approximation neglects all polarization effects on the propagation constants of the eigenmodes of the fiber, and yields a single propagation constant for a given value of $l$ \cite{SnyderA.W2012Owt}. In reality, for any given $l>2$ there exists a set of 4 modes, specifically OAM$_{\pm l,m}^{s}$ modes with different combinations of $\pm l$ and $s = \pm 1$. The modes with $\pm l$ and $s = \pm 1$ (same sign) are degenerate and called the spin-orbit aligned modes, and have slightly higher propagation constants than the degenerate spin-orbit anti-aligned modes with $\pm l$ and $s = \mp 1$ (different sign). 

\begin{figure}
    \centering
    \includegraphics[width=.48\textwidth]{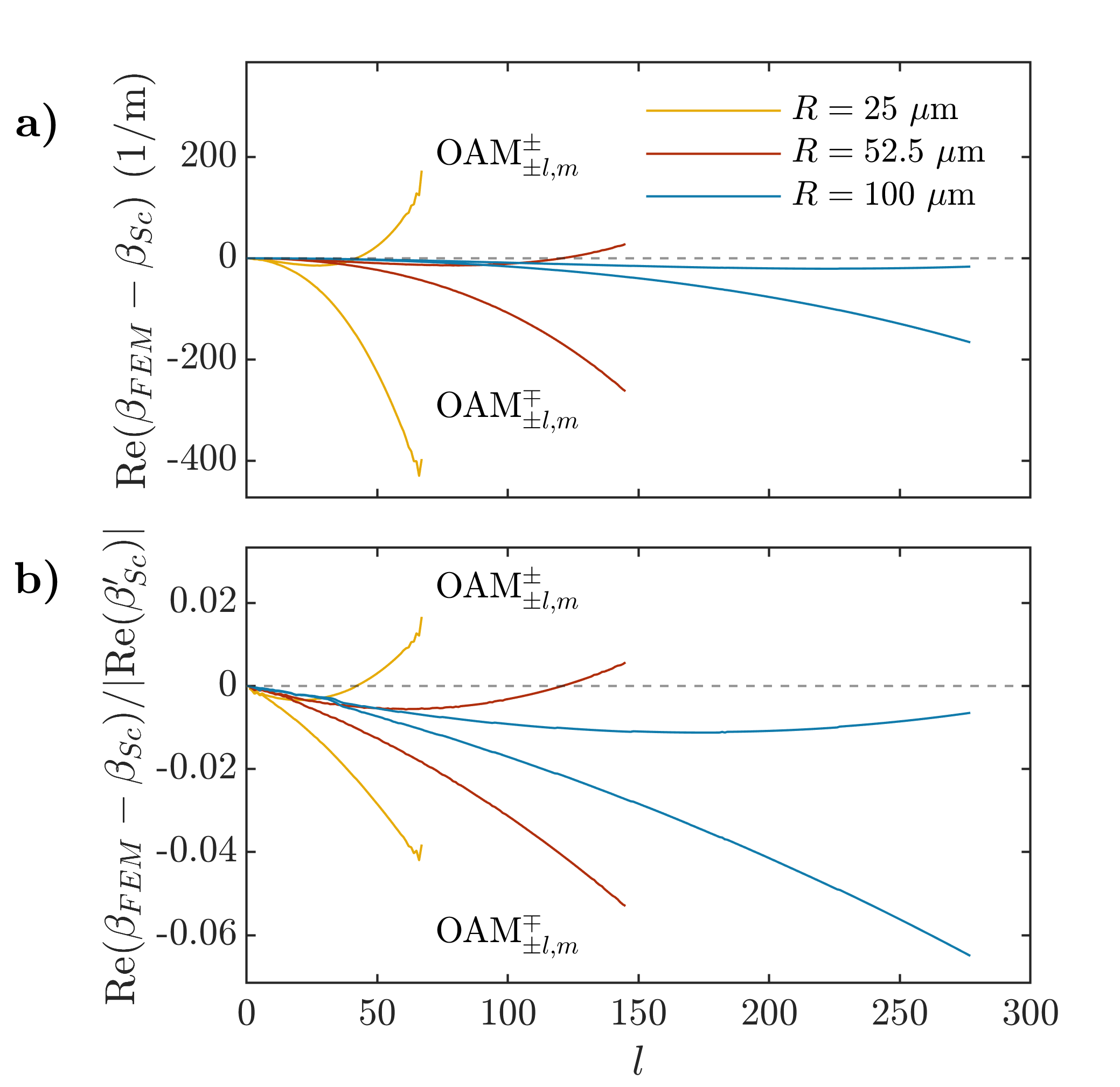}
    \caption{\textbf{a)} The error between the scalar propagation constant $\beta_{Sc}$ and the propagation constants of the spin-orbit aligned, OAM$_{\pm l,m}^{\pm}$ and spin-orbit anti-aligned OAM$_{\pm l,m}^{\mp}$ modes as retrieved via FEM simulations in COMSOL, within the region $\beta>0.98 k_{cl}$. \textbf{b)} The error relative to the magnitude of the first derivative of the real part of the scalar propagation constant $|\textrm{Re}(\beta_{Sc}')|$.}
    \label{fig_s4}
\end{figure}

In Fig. \ref{fig_s4} we plot the difference between the scalar propagation constant from Eq. 4 in the main text and propagation constants of the spin-orbit aligned and spin-orbit anti-aligned modes retrieved via FEM simulations in COMSOL, within the region $\beta>0.98 k_{\text{cl}}$. Already visible in the error of the modes of the $R = 25~\mu$m fiber, the error does not behave smoothly for higher $l$ (or lower Re$(\beta)$). The likely reason for this is that since higher-order modes leak more power from the core, the cladding structure becomes more relevant --- however the cladding structure is only considered in the FEM simulations, and not at all in the eigenvalue Eq. 4 in the main text that yields the scalar propagation constants. 

Nevertheless, the error of the weakly-guiding approximation is negligible in comparison to the strength of the modal dispersion (see Fig. 2b in the main text) relevant for the self-imaging effect. As seen in Fig. \ref{fig_s4}b, the error remains within a few percent of the first derivative of the real part of the scalar propagation constant Re($\beta_{Sc}'$) in the region of OAM we consider.

\bibliography{si}